\documentclass[12pt]{article}
\usepackage{graphicx}

\newcommand{\beq}{\begin{equation}}
\newcommand{\eeq}{\end{equation}}
\newcommand{\beqn}{\begin{eqnarray}}
\newcommand{\eeqn}{\end{eqnarray}}
\newcommand{\beqns}{\begin{eqnarray*}}
\newcommand{\eeqns}{\end{eqnarray*}}

\begin{document}

\begin{titlepage}
\begin{center}

\hfill USTC-ICTS-14-10\\
\hfill June 2014

\vspace{2.5cm}

{\large {\bf  A note on Higgs decays into $Z$ boson and
$J/\Psi (\Upsilon)$}}\\
\vspace*{1.0cm}
 {Dao-Neng Gao$^\dagger$ \vspace*{0.3cm} \\
{\it\small Interdisciplinary Center for Theoretical Study,
University of Science and Technology of China, Hefei, Anhui 230026
China}}

\vspace*{1cm}
\end{center}
\begin{abstract}
\noindent
Rare decays $h\to Z V$ with $V$ denoting the narrow $c\bar{c}$ or $b\bar{b}$ resonances, such as $J/\Psi$ or $\Upsilon$ states, have been analyzed. Within the standard model, these channels may proceed through the tree-level transition $h\to ZZ^*$ with the virtual $Z^*\to V$, and also loop-induced process $h\to Z\gamma^*$, followed by $\gamma^*\to V$. Our analysis shows that, for the bottomonium final states, the decay rate of $h\to Z \Upsilon$ from the loop-induced process is small and the former transition gives the dominant contribution; while, for the charmonium final states, $\Gamma(h\to Z J/\Psi)$ and $\Gamma(h\to Z\Psi(2S))$ induced by
$h\to Z\gamma^* \to Z V$ could be comparable to the contribution given by the tree-level $h\to ZZ^*\to Z V$ transition.
\end{abstract}

\vfill
\noindent
$^{\dagger}$ E-mail:~gaodn@ustc.edu.cn
\end{titlepage}

 After the discovery of the 125 GeV Higgs boson at the Large Hadron Collider (LHC) \cite{atlascms}, a new era of the precise determination of the properties of this new particle has begun. So far, current measurements of the Higgs boson couplings to standard model (SM) fields are consistent with those expected within the SM, it is nevertheless conceivable that more in-depth investigations both theoretically and experimentally, may reveal the non-standard properties of the particle. Of these studies, the so-called golden channel, $h\to Z Z^*\to 4 \ell$, might be an interesting decay mode towards accomplishing this goal, which is capable of both probing the nature of general $hZZ$ couplings including the CP properties \cite{hzz} and exploring exotic Higgs decays \cite{exotic, Isidori2014} that are not predicted by the SM.

In Ref. \cite{Isidori2014}, the $h\to 4 \ell$ decay spectrum has been analyzed in the kinematical region where low invariant mass of the dilepton pair, around several GeV,  are not far from QCD resonances (such as heavy quarkonia $J/\Psi$ and $\Upsilon$), and non-perturbative QCD effects near from these quarkonium thresholds, induced by $h\to Z V \to Z \ell^+\ell^-$ (where $V=J/\Psi$ or $\Upsilon$ etc.), has to be taken into account. In the present work, we will focus on the rare decays $h\to Z V$ themselves, instead of their contributions to the $h\to 4 \ell$ spectrum.

The decay rates of $h\to Z V$  have been calculated in Refs. \cite{Isidori2014,IMM2013} via the tree-level vertex $h\to Z Z^*$, with the subsequent transition $Z^*\to V$. The purpose of this note is to show that, these decays can also proceed through $h\to Z \gamma^*$, followed by $\gamma^*\to V$. Although in the SM, $h\to Z \gamma$ transition is loop-suppressed, our analysis below shall indicate that, due to the smallness of the masses of $V$'s, the amplitude of $h\to Z\gamma^*\to Z V$ could be enhanced, which may thus bring about significant contributions to these processes. Similar transition $h\to \gamma\gamma^*$ with $\gamma^*\to V$ has also been studied in $h\to \gamma J/\Psi (\Upsilon)$ decays by the authors of Ref. \cite{BPSV13}.

In the SM, the vertex of  Higgs coupling to $Z$ pair is
\beq\label{hzz}{\cal L}_{h ZZ}=\frac{m_Z^2}{v}h Z_\mu Z^\mu,
\eeq
and the neutral current interactions are expressed as
\beq\label{NC}
{\cal L}_{\rm NC}=e J_\mu^{\rm em} A^\mu+\frac{g}{\cos\theta_W}J^Z_\mu Z^\mu\eeq
with
\beq\label{emcurrent}
J_\mu^{\rm em}=\sum_f Q_f\bar{f}\gamma_\mu f,
\eeq
and
\beq\label{weakneutralcurrent}
J_\mu^Z=\frac{1}{2}\sum_f \bar{f}\gamma_\mu(g_V^f-g_A^f\gamma_5)f.\eeq
Here $v=(\sqrt{2}G_F)^{-1/2}\approx$ 246 GeV, $e$ is the QED coupling constant, $g$ is the SU(2)$_L$ coupling constant, $\theta_W$ is the Weinberg angle, and $f$ denotes fermions including leptons and quarks.  Also $g_V^f=T_3^f-2 Q_f \sin^2 \theta_W$ and $g_A^f=T_3^f$,  where $Q_f$ is the charge, and $T_3^f$ is the third component of the weak isospin of the fermion.

\begin{figure}[t]
\begin{center}
\includegraphics[width=7cm,height=3cm]{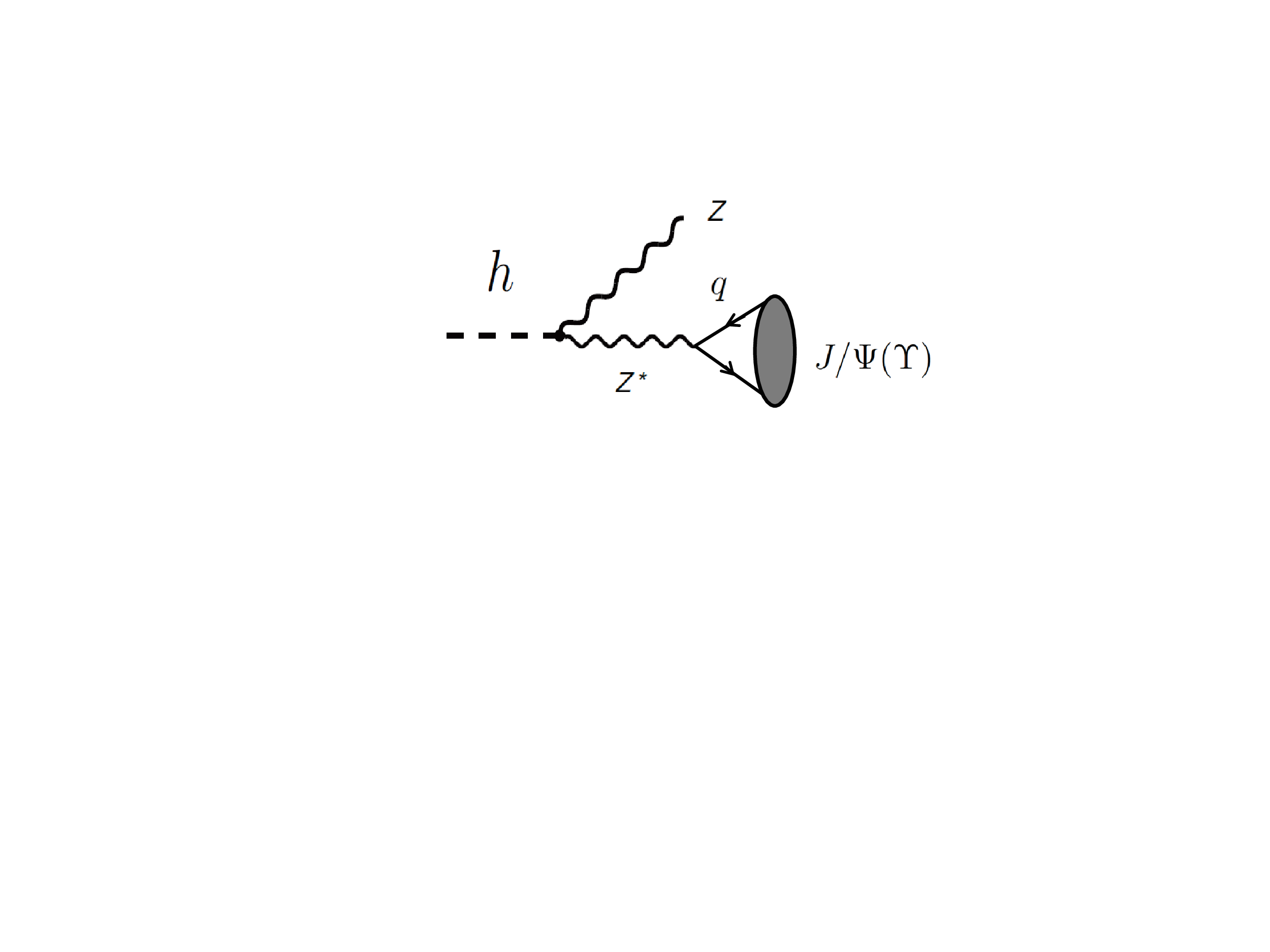}
\end{center}
\caption{The diagram for $h\to Z Z^*\to Z V$. The $h ZZ^*$ vertex is from Eq. (\ref{hzz}), and the virtual $Z$ boson coupling to quark pair is from  Eq. (\ref{NC}).}\label{figure1}
\end{figure}

The amplitude of $h\to Z V$ via $h\to ZZ^*\to Z V$, as shown in Fig. \ref{figure1}, has been calculated in Ref. \cite{Isidori2014}, which reads
\beq\label{amplitude1}
{\cal M}_1=-\frac{2 m_Z^2 g}{v\cos\theta_W}\frac{1}{m_Z^2-m_{V}^2}g_V^q f_{V}m_{V}\epsilon_{V} \cdot \epsilon_Z.\eeq
Here the superscript $q=c$ or $b$,  and the decay constant $f_{V}$ can be defined by \cite{Isidori2014, IMM2013}
\beq\label{fv}
\langle 0|\bar{q}\gamma^\mu q |V(p,\epsilon)\rangle= f_{V}m_{V}\epsilon_{V}^\mu, 
\eeq
where $\epsilon_V^\mu$ is the polarization vector for $c\bar{c}$ or  $b\bar{b}$ narrow resonances with $J^{PC}=1^{--}$.

Next let us discuss the amplitude intermediated by the virtual photon. In the SM,  the $h\to Z\gamma$ decay at the leading order is determined by the one-loop contribution \cite{hgammaz}. On the other hand, more explicitly, one can write down the effective lagrangian for the $h Z\gamma$ interactions generated in the SM as follows \cite{HS94, KK2013}
\beq\label{effectivehzgamma}
{\cal L}^{hZ \gamma}_{\rm eff}=\frac{e g}{16\pi^2 v} C_{Z\gamma} Z_{\mu \nu} F^{\mu\nu} h,\eeq
where $C_{Z\gamma}$ is dimensionless effective coupling constant. In the SM at the one-loop order, $C_{Z\gamma}$ will get contributions from $W$-boson and top-quark loop diagrams, whose explicit expressions can be found in Ref. \cite{GHKD90, KK2013}. For the general effective $h Z\gamma$ interactions beyond the SM, some new structures other than eq. (\ref{effectivehzgamma}) may of course appear \cite{HS94, KK2013}.

\begin{figure}[t]
\begin{center}
\includegraphics[width=7cm,height=3cm]{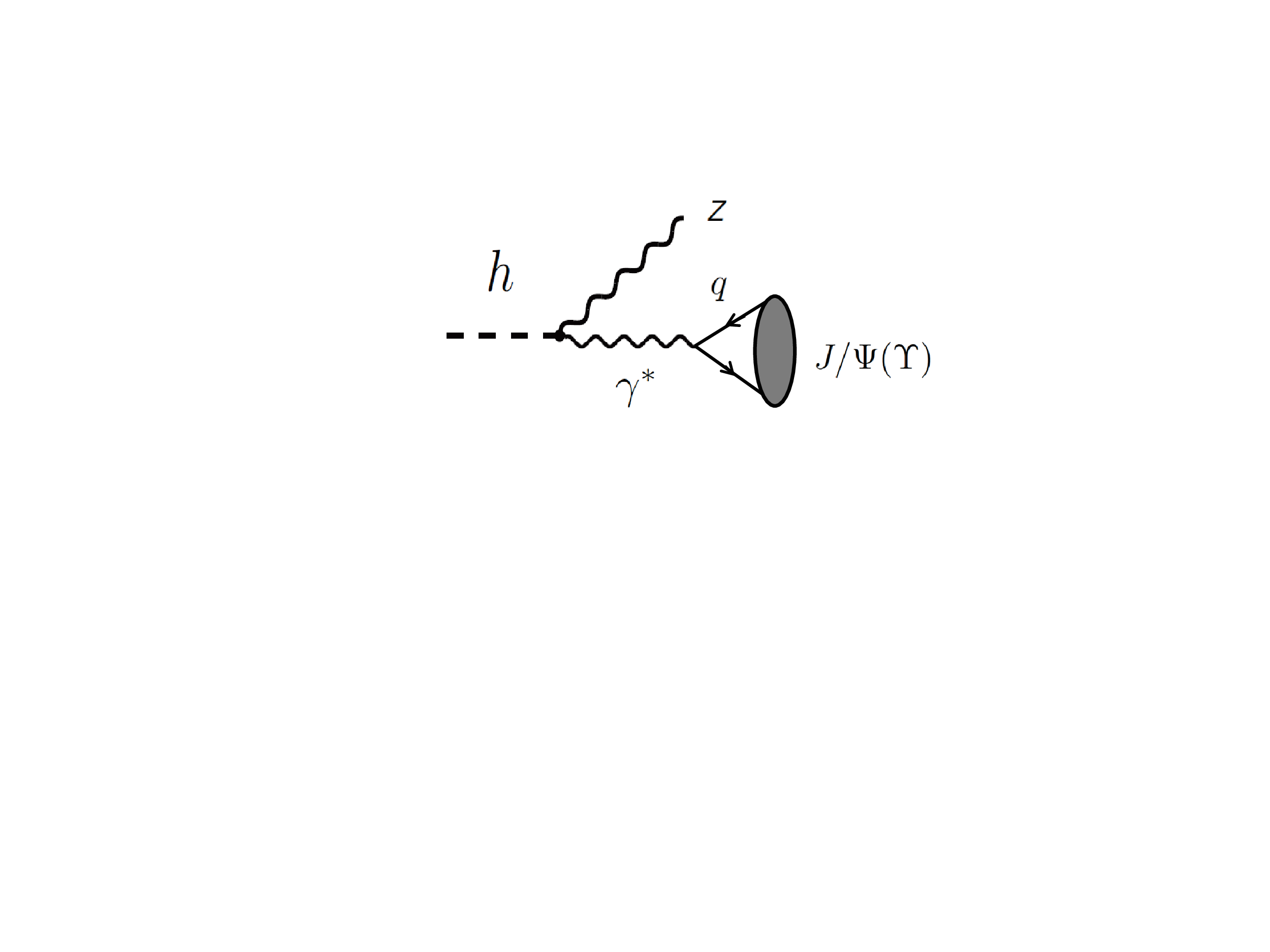}
\end{center}
\caption{The diagram for $h\to Z \gamma^*\to Z V$. The $h Z\gamma^*$ vertex is from the effective interaction in Eq. (\ref{effectivehzgamma}), and the virtual photon coupling to quark pair is from  Eq. (\ref{NC}).}\label{figure2}
\end{figure}

Now the amplitude of $h\to Z V$ through $h\to Z\gamma^*\to Z V$ transition, as shown in Fig. \ref{figure2}, can be given by
\beq\label{amplitude2}
{\cal M}_2=\frac{\alpha_{\rm em}g f_{V}Q_q}{2\pi v} \frac{C_{Z\gamma}}{m_{V}}(k_{\mu} p_\nu- k\cdot p  g_{\mu\nu}) \epsilon_Z^\nu\epsilon_{V}^\mu,\eeq
where $\alpha_{\rm em}=e^2/4\pi$. $k$ and $p$ are 4-momenta of $Z$-boson and the resonance $V$, respectively. It is seen that there is a $1/m_V$ factor in the above equation, which comes from the virtual photon propagator ($1/m_V^2$) and $m_V$ in the matrix element of eq. (\ref{fv}).  Also we have used $C_{Z\gamma}$ for the on-shell photon case instead of the off-shell $C_{Z\gamma^*}$, since
\beq\label{off-shell}C_{Z\gamma^*}=C_{Z\gamma}+O(m_{V}^2/m_h^2),\eeq
which should be a good approximation for our present purpose.

The total decay amplitude of $h\to Z V$ can thus be written as
\beq\label{totalamplitude} {\cal M}={\cal M}_1+{\cal M}_2.\eeq
After squaring the amplitude and summing the polarization of final particles, one will obtain the decay rate of the process
\beq\label{rate}
\Gamma(h\to Z V)=\Gamma_1+\Gamma_2+\Gamma_{12}\eeq
with
\beq\label{rate1}
\Gamma_1=\frac{m_h^3(g_V^q f_{V})^2 }{16\pi v^4}\frac{\lambda^{1/2}(1,r_Z, r_V)}{(1-r_V/r_Z)^2} [(1-r_Z-r_V)^2+8 r_Z r_V]\eeq
from ${\cal M}_1$,
\beq\label{rate2}\Gamma_2=\frac{\alpha_{\rm em}^3 f_{V}^2 Q_q^2m_h^3}{32\pi^2 v^2 \sin^2\theta_W}\frac{C_{Z\gamma}^2}{m_{V}^2}\lambda^{1/2}(1,r_Z, r_V)[(1-r_Z-r_V)^2+2 r_Z r_V]\eeq
from ${\cal M}_2$, and
\beq\label{rate12}
\Gamma_{12}=\frac{3\alpha_{\rm em}^2 f_{V}^2 g_V^q Q_q m_h C_{Z\gamma}}{8\pi \cos\theta_W \sin^2\theta_W v^2}\frac{\lambda^{1/2}(1,r_Z, r_V)}{1-r_V/r_Z}(1-r_Z-r_V)\eeq
from the interference between ${\cal M}_1$ and ${\cal M}_2$. Here $r_Z=m_Z^2/m_h^2$, $r_V=m_{V}^2/m_h^2$, and $\lambda(a,b,c)=a^2+b^2+c^2-2 (ab+ac+bc)$.

Using the decay width of Higgs boson $\Gamma_h\approx 4.07$ MeV, and defining
\beq\label{branchingratioi}
{\cal B}_i(h\to Z V)={\Gamma_i}/{\Gamma_h}
\eeq
for $i=1,~2$, and
\beq\label{branchingratio}
{\cal B}(h\to Z V)=\frac{\Gamma(h\to Z V)}{\Gamma_h},
\eeq
one can get branching ratios of $h\to Z V$ decays, which has been listed in Table 1.  ${\cal B}_1(h\to Z V)$ in the fourth column is given by the tree level transition $h\to ZZ^*\to Z V$ of Fig. \ref{figure1}, which has been calculated by the authors of Ref. \cite{Isidori2014}. The main results of this note are given in the fifth and sixth column. It is shown that, for the narrow $c\bar{c}$ resonances $J/\Psi(1S)$ and $\Psi(2S)$, contributions from Fig. \ref{figure2} via $h\to Z \gamma^*\to Z V$  (${\cal B}_2$) are significant; while for the bottomonium  resonances, it is a different story, and Fig. \ref{figure1} gives the dominant contribution. It is easy to see that the factor $1/m_V^2$ appearing in $\Gamma_2$ of eq. (\ref{rate2}) plays an important role for this point. Due to the smallness of $J/\Psi$ and $\Psi(2S)$ masses, it is obvious that $\Gamma_2$ for the charmonium resonances could be rather enhanced. Comparing with the $c\bar{c}$ case, however, the relative large masses of $b\bar{b}$ states and the electric charge of bottom quark ($Q_b=-1/3$, $Q_c=2/3$) will lead to a factor about $1/40$ suppression in $\Gamma_2$ if we do not take into account the difference of $f_V$'s.

\begin{table}[t]\begin{center}\begin{tabular}{ c|c| c| c| c| c} \hline\hline
 Resonance & $m_{V}$(GeV)& $f_{V}$(MeV) & ${\cal B}_1(h\to Z V)$ &${\cal B}_2(h\to Z V)$ & ${\cal B}(h\to Z V)$\\\hline
 $J/\Psi(1S)$ & 3.097& 405 &$1.7\times 10^{-6}$ &$1.0\times 10^{-6}$ & $3.2\times 10^{-6}$\\\hline
$\Psi(2S)$& 3.686 & 290 &$8.7\times 10^{-7}$ &$3.7\times 10^{-7}$ &$1.5\times 10^{-6}$\\\hline
$\Upsilon(1S)$& 9.460& 680& $1.6\times 10^{-5}$& $7.5\times 10^{-8}$ & $1.7\times 10^{-5}$\\\hline
$\Upsilon(2S)$& 10.02& 485 & $8.3\times 10^{-6}$&$3.4\times 10^{-8}$& $8.9\times 10^{-6}$\\\hline
$\Upsilon(3S)$&10.36& 420&$6.3\times 10^{-6}$&$2.4\times 10^{-8}$&$6.7\times 10^{-6}$\\\hline
\hline
\end{tabular}\caption{Branching ratios of $h\to Z V$ decays with $V$ denoting the narrow $c\bar{c}$ and $b\bar{b}$ resonances with $J^{PC}=1^{--}$. The decay constants $f_{V}$'s in the third column are taken from Ref. \cite{Isidori2014}, and the values of ${\cal B}_1$ in the fourth column agree with the results given in Ref. \cite{Isidori2014}.} \end{center}\end{table}

Experimentally, the decay channel $h\to Z\gamma$ has been studied by ATLAS \cite{atlas2013} and CMS \cite{cms2013} at LHC. Within the SM, this process is loop-induced, which thus is sensitive to physics beyond the SM \cite{BSM}.
The above analysis shows that rare decays $h\to Z J/\Psi$ and $h\to Z \Psi(2S)$ could get significant contributions from $h \to Z\gamma^*$ followed by $\gamma^* \to J/\Psi$ or $\Psi(2S)$ transitions.  Thus the future precise experimental studies of these rare decays may provide some complimentary information for the $h\to Z\gamma$ decay, both in and beyond the SM.

To summarize, in the SM, the rare decay modes $h\to Z V$ with $V=J/\Psi$ or $\Upsilon$ states may happen through two ways, one is $h\to ZZ^*\to Z V$, the other is  $h\to Z\gamma^*\to ZV$. These decay rates via the first way have been evaluated in Ref. \cite{Isidori2014}. In order to complete the analysis, we calculate both of them in the present paper. Our study indicates that, for the narrow $b\bar{b}$ resonances, the decay rates via the second way is small and the first way gives the dominant contribution; while, due to the enhancement by the small masses of charmonium resonances, the decay rates of $h\to ZJ/\Psi$ and $h\to Z \Psi(2S)$ through the second way could be comparable to the contributions induced by $h\to Z Z^*\to ZV$.

\vspace{0.5cm}
\section*{Acknowledgements}
This work was supported in part by the NSF of China under Grant Nos. 11075149 and 11235010.

\end{document}